\documentclass[%
 reprint,
 nofootinbib,
 amsmath,amssymb,
 aps,
]{revtex4-2}
\pdfoutput=1
\usepackage{graphicx}% Include figure files
\usepackage{dcolumn}% Align table columns on decimal point
\usepackage{bm}% bold math
\usepackage [dvipsnames] {xcolor}
\usepackage{hyperref}% add hypertext capabilities
\hypersetup
{
colorlinks=true,
linkcolor=blue,
filecolor=blue,
urlcolor=blue,
citecolor=blue,
}
\usepackage[mathlines]{lineno}% Enable numbering of text and display math
\begin{document}

\preprint{APS/123-QED}

\title{Probing Primordial black holes with the distortion of Stochastic Gravitational Wave Background}

\author{Mingqi Sun$^{1}$}
\author{Kai Liao$^{1}$}
 \email{E-mail:liaokai@whu.edu.cn}
\author{Xi-Long Fan$^{1}$}%
 \email{E-mail:xilong.fan@whu.edu.cn}

\affiliation{$^{1}$School of Physics and Technology, Wuhan University, Wuhan 430072, China}

\date{\today}% It is always \today, today,

\begin{abstract}
The stochastic gravitational-wave background (SGWB), arising from the incoherent superposition of numerous compact binary coalescences, serves as a powerful probe of both astrophysical populations and fundamental physics. In this work, we investigate the influence of gravitational lensing on the SGWB, focusing on primordial black holes (PBHs) as potential lenses. Assuming PBHs as dark matter candidates with a broad cosmic distribution, we show that their lensing optical depth can be significantly enhanced, producing pronounced effects with relative deviations at the $10^{-1}$ level. By systematically varying the PBH mass $M_{\mathrm{PBH}}$ and abundance $f_{\mathrm{PBH}}$, we demonstrate that the mass predominantly determines the frequency-dependent diffraction features of the spectrum, while the abundance primarily amplifies the overall lensing-induced deviation. Although the SGWB from binary black holes has yet to be observed, our analytical results provide theoretical insight into the possible imprint of lensing on its spectrum and suggest that future detections could offer a novel avenue to constrain dark matter scenarios.
\end{abstract}

\maketitle

%\tableofcontents

\section{Introduction}
Dark matter, a hypothetical form of matter that constitutes approximately five times the mass density of ordinary baryonic matter on cosmological scales, remains one of the most profound mysteries in modern physics. Among various proposed candidates, primordial black holes (PBHs) have attracted significant attention as plausible dark matter constituents~\cite{bertone2018history}. A wide range of observational approaches have been developed to constrain the PBH abundance across different mass regimes~\cite{sasaki2018primordial,green2021primordial}, which are theorized to span from $10^{-18}$ to $10^{17}\,M_{\odot}$~\cite{frampton2016primordial}. In recent years, several promising observational probes have been proposed to further explore PBHs in forthcoming observations, including gravitational lensing of gravitational waves~\cite{liao2020probing,urrutia2022lensing}, gamma-ray bursts~\cite{ji2018strong}, and 21 cm cosmological signals~\cite{halder2021bounds}.

The stochastic gravitational-wave background (SGWB) arises from the incoherent superposition of a large number of independent gravitational-wave (GW) sources and can have both cosmological~\cite{grishchuk1977gravitational} and astrophysical~\cite{regimbau2011astrophysical,2025ChPhL..42e9201Z} origins. For a cosmological SGWB, the signal may be generated by quantum vacuum fluctuations in the early Universe, while for an astrophysical SGWB, the cumulative emission from compact binary coalescences, such as binary black hole (BBH) mergers, can contribute significantly. According to Ref.~\cite{abbott2018kagra}, the BBH-generated SGWB is expected to dominate within the LIGO--Virgo frequency band. The astrophysical SGWB is of particular interest for at least two reasons~\cite{zhu2011stochastic}. First, the spectral shape encodes global statistical information about the underlying source population, allowing one to infer properties such as merger rates and mass distributions. Second, it may serve as a foreground that obscures the detection of the cosmological SGWB originating from the early Universe.

In addition to GWs themselves, gravitational lensing is another fundamental prediction of general relativity, describing the deflection of light or radiation as it passes near massive objects. This phenomenon has been widely exploited in astrophysics to probe dark matter distributions, measure cosmological parameters, and test general relativity. Importantly, gravitational lensing can also affect GW signals~\cite{oguri2018effect,2022ChPhL..39k9801L,2017NatCo...8.1148L}, introducing characteristic distortions that depend on the mass and configuration of the lens. Two distinct regimes are generally considered in GW lensing. When the GW wavelength is much shorter than the Schwarzschild radius associated with the lens, the system can be well approximated in the geometric-optics limit, similar to electromagnetic lensing by galaxies or clusters. In contrast, when the GW wavelength is comparable to or exceeds the characteristic lens scale, diffraction effects become non-negligible, and the wave-optics framework must be adopted~\cite{bliokh1975diffraction}. The latter regime can introduce frequency-dependent modulations to the waveform, a phenomenon that has been extensively explored in the literature~\cite{1998PhRvL..80.1138N,takahashi2003wave,takahashi2017arrival,liao2019wave}.

The impact of gravitational lensing has already been extensively investigated in the context of the cosmic microwave background (CMB)~\cite{hanson2010weak}. However, despite the SGWB generated by BBH mergers being a promising probe of both astrophysical populations and early-Universe physics, the influence of lensing on the SGWB remains relatively unexplored. Motivated by this gap, the present study aims to develop an analytical framework to quantify the potential modifications to the SGWB spectrum induced by gravitational lensing.

Building upon the above considerations, we focus on the SGWB under the influence of PBH lensing and examine the resulting changes in its power spectrum. The distinctive lensing-induced features identified in this study may offer a novel means of constraining PBH properties once the SGWB is experimentally observed.

The remainder of this paper is organized as follows. In Sec.~2, we derive the fundamental analytical expression for the astrophysical SGWB produced by BBH mergers. In Sec.~3, we briefly review the wave-optics treatment of gravitational lensing. In Sec.~4, we construct the analytical model for the lensed SGWB, incorporating the relevant physical properties of both the lenses and the background, and derive the key expressions. Finally, in Sec.~5, we summarize our conclusions and discuss the broader implications of our results.
\section{Basic form of SGWB}
In this chapter, we present a complete analytical formulation of the stochastic gravitational wave background (SGWB) and explain the physical meaning of each component. In addition, given that numerous studies have adopted various parameter choices, we provide a detailed account of the parameter settings used in this work. 

First of all, we use the usual thinking to define the physical quantity, $\Omega_{\mathrm{GW}}(f)$ to describe the energy spectrum of gravitational wave, its equation of definition is:
\begin{equation}
    \Omega_{\mathrm{GW}}=\frac{1}{\rho_{c,0}}\frac{d\,\rho_{\mathrm{GW}}}{d\,lnf},
\end{equation}
where the $\rho_{c,0}$ is the critical energy density required to close the universe today, $\rho_{c,0}=3H_{0}^2c^2/8\pi G$ with $H_0$ the Hubble constant today, and the $\rho_{\mathrm{GW}}$ denotes the energy density of gravitational wave.

We begin with Phinney’s practical theorem \cite{phinney2001practical}, which provides a general analytical expression. To ensure that each component of the expression has a clear physical correspondence, we adopt a more explicit formulation presented in \cite{zhu2013gravitational}:
\begin{equation}
    \label{gwb}
    \Omega_{\mathrm{GW}}=\frac{f}{\rho_{c,0}H_{0}}\int^{z_{\mathrm{max}}}_{z_{\mathrm{min}}}\frac{R_{\mathrm{V}}(z_{s})}{(1+z_{s})E(z_{s})} \left.\left(\frac{dE_{\mathrm{GW}}}{df_r} \right)\right|_{f_r=(1+z_{s})f}dz_{s}.
\end{equation}
Here, \( E(z_{s}) = H(z_{s})/H_0 = \sqrt{\Omega_{\Lambda} + \Omega_{\mathrm{m}}(1 + z_{s})^3} \). In this work, we adopt the standard \( \Lambda \)CDM cosmological model with parameters \( H_0 = 67.66\,\mathrm{km\,s^{-1}\,Mpc^{-1}} \) and \( \Omega_{\mathrm{m}} = 0.31 \) \cite{aghanim2020planck}. The quantities \( R_{\mathrm{V}} \) and \( dE_{\mathrm{GW}}/df_r \) play a crucial role in this study, and we will discuss them in more detail in subsequent sections.

To be noticed, there are several different form of this equation, but they all have the same physical meaning.
\subsection{Specific form of BBH merge rate}
The physical meaning of \( R_{\mathrm{V}} \) is the number of events per unit cosmic time per unit comoving volume, where each event corresponds to the coalescence of a binary black hole system. To compute this important quantity, it is convenient to adopt a specific form \cite{regimbau2008astrophysical,wu2012accessibility}:
\begin{equation}
    \label{rv}
    R_{\mathrm{V}}(z) = \lambda\,\dot\rho_*(z)
\end{equation}
where \(\lambda\) denotes the mass fraction of the stars that end up in compact binary systems (with unit \(M_{\odot}^{-1}\)) and \( \dot\rho_*(z) \) represents the cosmic star formation rate density (SFR)(in units of \( M_{\odot}\,\mathrm{yr^{-1}\,Mpc^{-3}} \)). In practice, the value of \( \lambda \) varies depending on the detector sensitivity and assumed population models.

However, Eq.~\ref{rv} does not account for the time delay between the formation of a binary system (when the star formation rate applies) and the actual coalescence event observable through gravitational waves. To incorporate this effect—which is not yet fully understood despite many studies of the stochastic gravitational wave background (SGWB)—we modify the star formation rate density as follows:
\begin{equation}
    \label{rv_delay}
    \dot\rho_{*,d}(z)=\int_{t_{\min}}^{t_{\max}} \dot\rho_*\big(t(z)+t_d\big)\,P(t_d)\, dt_d,
\end{equation}
In this work, we adopt the lookback time as the time coordinate, defined as
\[
t(z) = \int_0^z \frac{1}{(1+z')H(z')} \, dz'.
\]
The function \( P(t_d) \) denotes the probability distribution of the time delay \( t_d \), which follows an inverse distribution motivated by studies of compact binary evolution \cite{dominik2012double}. The integration limits \( t_{\min} \) and \( t_{\max} \) specify the range of allowed time delays for the chosen distribution; we will discuss their specific values later. The corresponding event rate is then given by:
\begin{equation}
    \label{rv_time_delay}
    R_{\mathrm{V}}(z)=\lambda\,\dot\rho_{*,d}(z).
\end{equation}

\begin{figure}
    \centering
    \begin{minipage}{\linewidth}
        \centering
        \includegraphics[width=\linewidth]{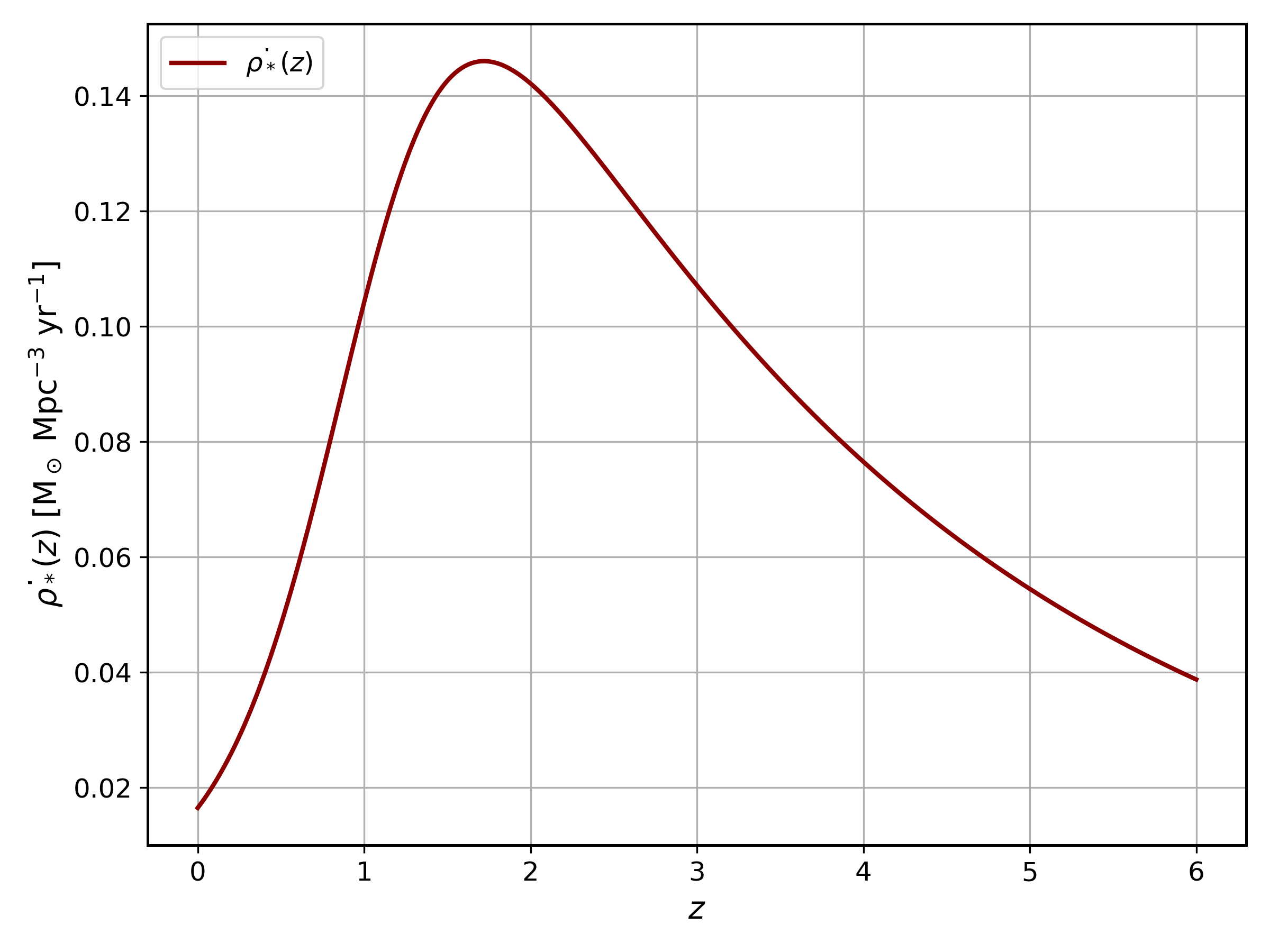}    
    \end{minipage}
    \caption{The normalized star formation rate (SFR) density. This curve illustrates the evolutionary trend of the BBH merger rate, denoted as \( R_{\mathrm{V}} \) in this work.}
    \label{RV}
\end{figure}

The final part of this subsection is to specify the functional form of the SFR, which is essential for computing the stochastic gravitational wave background (SGWB). Here, we adopt the parametrization used in the LIGO/Virgo analysis \cite{vangioni2015impact}:
\begin{equation}
    \label{SFR}
    \dot\rho_*(z)=\nu\,\frac{a\,\exp[b(z-z_{\mathrm{m}})]}{a-b + b\,\exp[a(z-z_{\mathrm{m}})]},
\end{equation}
where \( \nu = 0.146\,M_{\odot}\,\mathrm{yr^{-1}\,Mpc^{-3}} \), \( a = 2.80 \), \( b = 2.46 \), and \( z_{\mathrm{m}} = 1.72 \), the corresponding curve is shown in Fig.~\ref{RV}. As mentioned above, we introduce a normalization factor \( \lambda \) and calibrate it using Eq.~\ref{rv_time_delay}. Specifically, we adopt \( \lambda\,\nu = 23.9\,\mathrm{yr^{-1}\,Gpc^{-3}} \), based on the main results of the LIGO-Virgo GW Transient Catalog \cite{abbott2021population}.

In this analysis, we set the time delay interval to \( t_{\mathrm{min}} = 50\,\mathrm{Myr} \) \cite{dominik2012double} and \( t_{\mathrm{max}} = 10^4\,\mathrm{Myr} \), corresponding to the lookback time at redshift \( z_* = 6 \). The choice of \( t_{\mathrm{max}} \) is motivated by previous studies \cite{hopkins2006normalization, fardal2007evolutionary, wilkins2008evolution}, which indicate that the star formation rate becomes negligible at redshifts greater than 6.
\subsection{Energy spectrum of a single BBH}
In the Newtonian limit, the gravitational-wave (GW) energy spectrum of a circular binary with component masses \( m_1 \) and \( m_2 \) is given by \cite{thorne1997gravitational}:
\begin{equation}
    \label{ins}
    \frac{dE_{\mathrm{GW}}}{df_r} = \frac{(\pi G)^{2/3} M_c^{5/3}}{3} f_r^{-1/3},
\end{equation}
where \( M_c \) is the chirp mass defined as
\begin{equation}
M_c = \frac{(m_1 m_2)^{3/5}}{(m_1 + m_2)^{1/5}}.
\end{equation}
This expression describes the GW emission during the inspiral phase only. To construct a more complete waveform model, we adopt the inspiral-merger-ringdown (IMR) energy spectrum \cite{ajith2008template}:
\begin{equation}
    \label{imr}
    \frac{dE_{\mathrm{GW}}}{df_r} = \frac{(\pi G)^{2/3} M_c^{5/3}}{3}
    \left\{
    \begin{array}{ll}
        f_r^{-1/3}, & f_r < f_1, \\
        \omega_1 f_r^{2/3}, & f_1 \leq f_r < f_2, \\
        \omega_2 \left[ \frac{f_r}{1 + \left( \frac{f_r - f_2}{\sigma/2} \right)^2} \right]^2, & f_2 \leq f_r < f_3,
    \end{array}
    \right.
\end{equation}
where \( \omega_1 = f_1^{-1} \) and \( \omega_2 = f_1^{-1} f_2^{-4/3} \) are normalization constants chosen to ensure continuity of \( dE_{\mathrm{GW}}/df_r \) at the transition frequencies \( f_1 \) and \( f_2 \).

The transition frequencies \( f_1 \), \( f_2 \), \( f_3 \), and the width parameter \( \sigma \), are determined by the total mass \( M = m_1 + m_2 \) and the symmetric mass ratio \( \eta = m_1 m_2 / M^2 \). These parameters can be computed using the fitting formula:
\begin{equation}
\label{IMR}
f = \frac{a_{k} \eta^2 + b_{k} \eta + c_{k}}{\pi M},
\end{equation}
where the coefficients \( a_{k} \), \( b_{k} \), and \( c_{k} \) are provided in Table \ref{tab:coefficients}.
\begin{table}[b]
\caption{\label{tab:coefficients}
Coefficients $a_k$, $b_k$, and $c_k$ for the fitting formula \ref{IMR}\cite{ajith2008template}.}
\begin{ruledtabular}
\begin{tabular}{lccc}
\textrm{Parameter} &
\textrm{$a_k$} &
\textrm{$b_k$} &
\textrm{$c_k$} \\
\colrule
$f_{\mathrm{1}}$ & $2.9740\times10^{-1}$ & $4.4810\times10^{-2}$ & $9.5560\times10^{-2}$ \\
$f_{\mathrm{2}}$ & $5.9411\times10^{-1}$ & $8.9794\times10^{-2}$ & $1.9111\times10^{-1}$ \\
$\sigma$            & $5.0801\times10^{-1}$ & $7.7515\times10^{-2}$ & $2.2369\times10^{-2}$ \\
$f_{\mathrm{3}}$  & $8.4845\times10^{-1}$ & $1.2848\times10^{-1}$ & $2.7299\times10^{-1}$ \\
\end{tabular}
\end{ruledtabular}
\end{table}

In this work, we assume that all binary black holes have equal component masses of \( 10\,M_{\odot} \), leading to a chirp mass of \( M_c = 8.7\,M_{\odot} \). Under this assumption, the characteristic frequencies are evaluated as:
\begin{equation}
    \label{para}
    (f_1, f_2, \sigma, f_3) = (404,\, 807,\, 237,\, 1153)\,\mathrm{Hz}.
\end{equation}
\section{Wave optics description}
In order to study the wave effect of gravitation lensing, we consider the weak gravitational field of the lens, so that the metric can be written as:
\begin{equation}
    \label{metric}
    ds^2=-(1+2U)dt^2+(1-2U)d\mathbf{r}^2,
\end{equation}
where \(U(r)\ll1\) is the gravitational potential of the lens object. Then we consider the linear perturbation \(h_{\mu\nu}\) in the background metric of lens object:
\begin{equation}
    \label{per}
    g_{\mu\nu}=g_{\mu\nu}^{(\mathrm{B})}+h_{\mu\nu},
\end{equation}
and if the wavelength \(\lambda\) is much smaller than the typical radius of the curvature of the background, we have \(h_{\mu\nu\,;\alpha}^{\quad\,\,\,\,;\alpha}=0\).
Following the eikonal approximation\cite{baraldo1999gravitationally}, we can express the gravitational wave as a scalar wave instead of a tensor wave: \(h_{\mu\nu}=\phi \,e_{\mu\nu}\). The propagation equation can be written as :
\begin{equation}
    \label{propagation}
    \partial_{\mu}\left(\sqrt{-g^{\mathrm{(B)}}}\,g^{\mathrm{(B)\mu\nu}}\,\partial_{\nu}\phi\right)=0.
\end{equation}
For the frequency domain \(\tilde{\phi}(f,\mathbf{r})\), the equation above can be rewritten as a Helmholtz equation:
\begin{equation}
    \label{ham}
    (\nabla^2+4\pi^2f^2)\tilde{\phi}=16\pi^2f^2U\tilde{\phi}.
\end{equation}
And it's convenient to define a amplitude factor, \(F(f)=\tilde{\phi}^{\mathrm{L}}(f)/\tilde{\phi}(f)\), where \(\tilde{\phi}^{\mathrm{L}}(f)\) and \(\tilde{\phi}(f)\) are the lensed and unlensed (\(U=0\)) amplitudes, respectively.

By solving the Eq.~\ref{ham}, the amplitude factor can be represent as :
\begin{equation}
    \label{amplitude}
    F(f)=\frac{D_{\mathrm{s}}R_{\mathrm{E}}^2(1+z_{l})}{D_{\mathrm{l}}D_{\mathrm{ls}}}\frac{f}{i}\int d^2\mathbf{x}\exp[2\pi i ft_{d}(\mathbf{x},\mathbf{y})],
\end{equation}
where \(R_{\mathrm{E}}\) denotes the Einstein radius of the lens while \(\mathrm{x}=\xi/R_{\mathrm{E}}\,,\,\mathbf{y}=\eta D_{\mathrm{l}}/(R_{\mathrm{E}}D_{\mathrm{s}})
\) which denote the dimensionless impact parameter in the lens plane and source position. The arrival time \(t_d\) at the observer from the source position \(\eta\) through \(\xi\) is given by\cite{takahashi2003wave}:
\begin{equation}
    \label{wave}
    t_d(\mathbf{x}\,,\,\mathbf{y})=\frac{D_{\mathrm{s}}R_{\mathrm{E}}^2(1+z_{l})}{D_{\mathrm{l}}D_{\mathrm{ls}}}\left[\frac{1}{2}|\mathbf{x-y}|^2-\psi(\mathbf{x})+\phi_{m}(\mathbf{y})\right],
\end{equation}
where \(\psi(\mathbf{x})\) is the dimensionless deflection potential, it's determined by the profile of the lens and \(\phi_{m}(\mathbf{y})\) is chosen so that the minimum arrival time is zero. Throughout this section, we adopt natural units by setting \( c = G = 1 \). For key equations such as Eq.~\ref{amplitude}\,,\ref{wave}, restoring physical units requires reintroducing a factor of \( c \) in the denominator to recover the SI unit system.

\section{Mechanisms of Gravitational Lensing on the SGWB}
In this section we will talk about the effect of gravitational lensing on the SGWB. We will give a analytical form of the lensed SGWB, and give explanation of how we can build this form. And we will also use it to calculate the effect of primordial black hole (as a candidate of dark matter) lensing on the SGWB.

It is well established that the SGWB arises from the incoherent superposition of a large number of individual gravitational wave events occurring throughout the Universe. Therefore, in order to understand how gravitational lensing affects the energy spectrum of the SGWB, it is essential to first examine how lensing modifies the energy spectrum of a single gravitational wave event. For a single event, the relation between gravitational wave caused by it and its energy spectrum is as follows:
\begin{figure}
    \centering
    \includegraphics[width=0.7\linewidth]{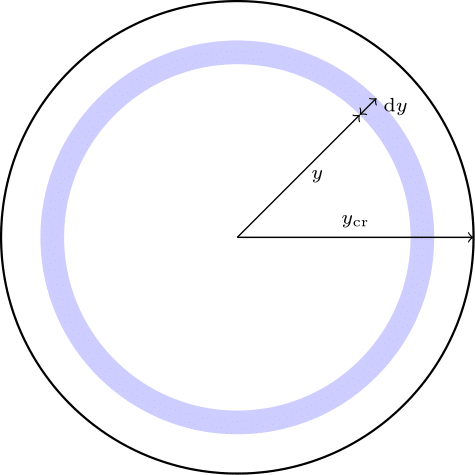}
    \caption{Monte Carlo sampling of source positions within a circle. The probability that a point falls within an annular ring between $y$ and $y + dy$ is proportional to the area of the ring, i.e., $2\pi y\,dy$. Therefore, the normalized probability density function is given by Eq.~\ref{monte}.}
    \label{montefig}
\end{figure}
\begin{equation}
    \label{spe}
    \frac{dE_{\mathrm{GW}}}{df}=\frac{2\pi^{2}c^{3}}{G}d_{\mathrm{M}}^2f^2\left \langle|\tilde{h}_{+}|^2+|\tilde{h}_{\times}|^2\right \rangle_{\Omega_{\mathrm{s}}},
\end{equation}
where \(d_{\mathrm{M}}=d_{\mathrm{L}}/(1+z)\) is so called proper-motion distance, \(\tilde{h}_{+,\times}(f)\) is waveform in frequency domain. So the relation between lensed and unlensed waveform is as follows:
\begin{equation}
    \label{lensedwavform}
    \tilde{h}^{\mathrm{L}}_{+,\times}(f)=F(f)\tilde{h}_{+,\times}(f).
\end{equation}
So if we assume the amplitude factor, \(F(f)\), is isotropy, we can than get the lensed energy spectrum for a single event:
\begin{equation}
    \label{lensedsingle}
    \left(\frac{dE_{\mathrm{GW}}}{df}\right)^{\mathrm{L}}=|F(f)|^2\,\frac{dE_{\mathrm{GW}}}{df}.
\end{equation}

In addition, it is necessary to estimate the fraction of events that can be gravitationally lensed. To this end, we introduce the concept of lensing optical depth, \(\tau(z_{s})\), a physical quantity that characterizes the probability of a given astrophysical event being lensed. This concept has been widely adopted in the literature to quantify lensing probabilities (e.g., \cite{takahashi2003wave,zhou2022search}). Certainly, the analytical form of the optical depth depends on the specific lens model under consideration. Different lens models require distinct formulations, which must be carefully treated. We will discuss these in detail in subsequent sections of this work.

Now we have all the tools we need to build analytical form of lensed SGWB:
\begin{widetext}
\begin{align}
\label{lensedsgwb}
\Omega_{\mathrm{GW}}^{\mathrm{L}}(f) = \frac{f}{\rho_{c,0}H_0} \Bigg[ &
\int_{z_{\mathrm{min}}}^{z_{\mathrm{max}}} \frac{R_{\mathrm{V}}(z_{s})}{(1+z_{s})\,E(z_{s})}\,\tau(z_{s})
\left.\left( \frac{dE_{\mathrm{GW}}}{df_r} \right)\right|_{f_r=(1+z_{s})f}
\left.\left\langle \left|F(f_l)\right|^2 \right\rangle\right|_{f_l=(1+z_{l}^*)f}\,dz_{s} \nonumber \\
& + \int_{z_{\mathrm{min}}}^{z_{\mathrm{max}}} \frac{R_{\mathrm{V}}(z_{s})}{(1+z_{s})\,E(z_{s})}\,(1 - \tau(z_{s}))
\left.\left( \frac{dE_{\mathrm{GW}}}{df_r} \right)\right|_{f_r=(1+z_{s})f}\,dz_{s} \Bigg].
\end{align}
\end{widetext}
In realistic scenarios, each individual gravitational wave event, if lensed, would be affected by a lens located at a different redshift and characterized by distinct lens parameters such as mass and position. However, in order to derive analytical expressions, we adopt a simplifying assumption that all events are lensed by identical lenses, with fixed redshift and mass (and other relevant parameters). As a result, the analytical expressions presented above do not incorporate any distribution over lens properties. Naturally, future studies may relax this assumption by introducing lens parameter distributions to improve the accuracy of the model.

No matter the exact functional form of \(\tau(z_s)\), it can always be expressed as an integral:$\tau(z_s) = \int_0^{z_s} f(z_l)\,dz_l,$ where \(f(z_l)\) can be interpreted as the probability density function (PDF) of the lensing rate. Consequently, the lens redshift \(z_l^*\) can be defined as the most probable redshift, determined by the condition:
\begin{equation}
    \label{defpro}
    \left.\frac{d f(z_l)}{dz_l}\right|_{z_l = z_l^*} = 0.
\end{equation}
An additional question concerns the choice of \(y\), which represents the source position in each lensing system. In this work, we adopt a Monte Carlo method by randomly sampling points within a circle of radius \(y_{\text{cr}}\), see in Fig.~\ref{montefig}. The radial distance of each point from the center is taken as the value of \(y\), leading to the following probability density function:
\begin{equation}
    \label{monte}
    P_d(y) = \frac{2y}{y_{\text{cr}}^2},
\end{equation}
where \(y_{\text{cr}}\) denotes the critical source position. Physically, if \(y < y_{\text{cr}}\), the corresponding event is considered to be lensed. Based on this distribution, the average value of the amplification factor is given by:
\begin{equation}
    \label{averageamplitudefactor}
    \left\langle \left|F(f_l)\right|^2 \right\rangle\Big|_{f_l = (1 + z_l^*)f}
    = \int_0^{y_{\text{cr}}} \left|F(f_l)\right|^2\Big|_{f_l = (1 + z_l^*)f} \frac{2y}{y_{\text{cr}}^2}\,dy.
\end{equation}

According to Takahashi \& Nakamura (2003)~\cite{takahashi2003wave}, the optical depth \(\tau\) can exceed unity, implying that a single event may be lensed multiple times. Consequently, Eq.~\ref{lensedsgwb} needs to be modified to account for this multiplicity:
\begin{widetext}
\begin{align}
\label{lensedsgwb_modify}
\Omega_{\mathrm{GW}}^{\mathrm{L}}(f) = \frac{f}{\rho_{c,0} H_0} \Bigg\{ &
\int_{z_{\mathrm{min}}}^{z_{\mathrm{max}}} \frac{R_{\mathrm{V}}(z_s)}{(1+z_s) E(z_s)} \left[1 - \lceil \tau(z_s) \rceil + \tau(z_s)\right]
\left.\left( \frac{dE_{\mathrm{GW}}}{df_r} \right)\right|_{f_r = (1 + z_s)f}
\left[\left\langle \left|F(f_l)\right|^2 \right\rangle\Big|_{f_l = (1 + z_l^*)f} \right]^{\lceil \tau(z_s) \rceil}\,dz_s \nonumber \\
& + \int_{z_{\mathrm{min}}}^{z_{\mathrm{max}}} \frac{R_{\mathrm{V}}(z_s)}{(1+z_s) E(z_s)} \left[\lceil \tau(z_s) \rceil - \tau(z_s)\right]
\left.\left( \frac{dE_{\mathrm{GW}}}{df_r} \right)\right|_{f_r = (1 + z_s)f}
\left[\left\langle \left|F(f_l)\right|^2 \right\rangle\Big|_{f_l = (1 + z_l^*)f} \right]^{\lfloor \tau(z_s) \rfloor}\,dz_s \Bigg\},
\end{align}
\end{widetext}
where \(\lceil \cdot \rceil\) and \(\lfloor \cdot \rfloor\) denote the ceiling and floor functions, respectively.

To quantify the difference between the lensed and unlensed energy spectra, we define the relative quantity
\begin{equation}
    \label{delta_f}
    \Delta(f)=\frac{\Omega^{\mathrm{L}}(f)-\Omega^{\mathrm{UnL}}(f)}{\Omega^{\mathrm{UnL}}(f)}.
\end{equation}

Several studies have investigated the lensing probability of point-mass lenses, e.g., \cite{zhou2022search,takahashi2003wave}. In this work, we consider primordial black holes (PBHs) as the lensing objects, since they are a viable dark matter candidate and may be distributed throughout the Universe. According to \cite{liao2019wave}, the lensing effect becomes significant when $y \leq 3$, in contrast to the standard strong-lensing scenario, where $y_{\mathrm{cr}}=1$ is typically assumed. This feature arises from gravitational-wave diffraction effects. The optical depth for a GW source at redshift $z_s$ is then given by:
\begin{equation}
    \label{optical_depth_point_mass}
    \tau(z_s)=\frac{3}{2}f_{\mathrm{PBH}}\Omega_{\mathrm{DM}}\int_{0}^{z_s}dz_l\,\frac{H_0^2}{cH(z_l)}\frac{D_{\mathrm{l}}D_{\mathrm{ls}}}{D_{\mathrm{s}}}(1+z_l)^2y_{\mathrm{cr}}^2,
\end{equation}
where we set $y_{\mathrm{cr}}=3$. Astrophysical black holes may also contribute to the lensed SGWB, $ \Omega^{\mathrm{L}}_{\mathrm{GW}} $. In this work, however, we focus on the upper-limit impact of compact-object lensing and therefore neglect the astrophysical black-hole contribution. The PBH abundance as a dark matter component is still uncertain, but for a not very small PBH fraction $ f_{\mathrm{PBH}} $, the lensing probability can be substantially larger than that of astrophysical black holes. Even under the extreme assumption that all baryonic matter forms black holes, their total abundance would be only $\sim 0.04 $, compared to a possible PBH fraction of $\sim 0.26$ (assuming total matter density $\Omega_M=0.3$). In the future, one may distinguish them with the knowledge of stars or according to their mass distributions. 

\begin{figure}
    \centering
    \includegraphics[width=1.0\linewidth]{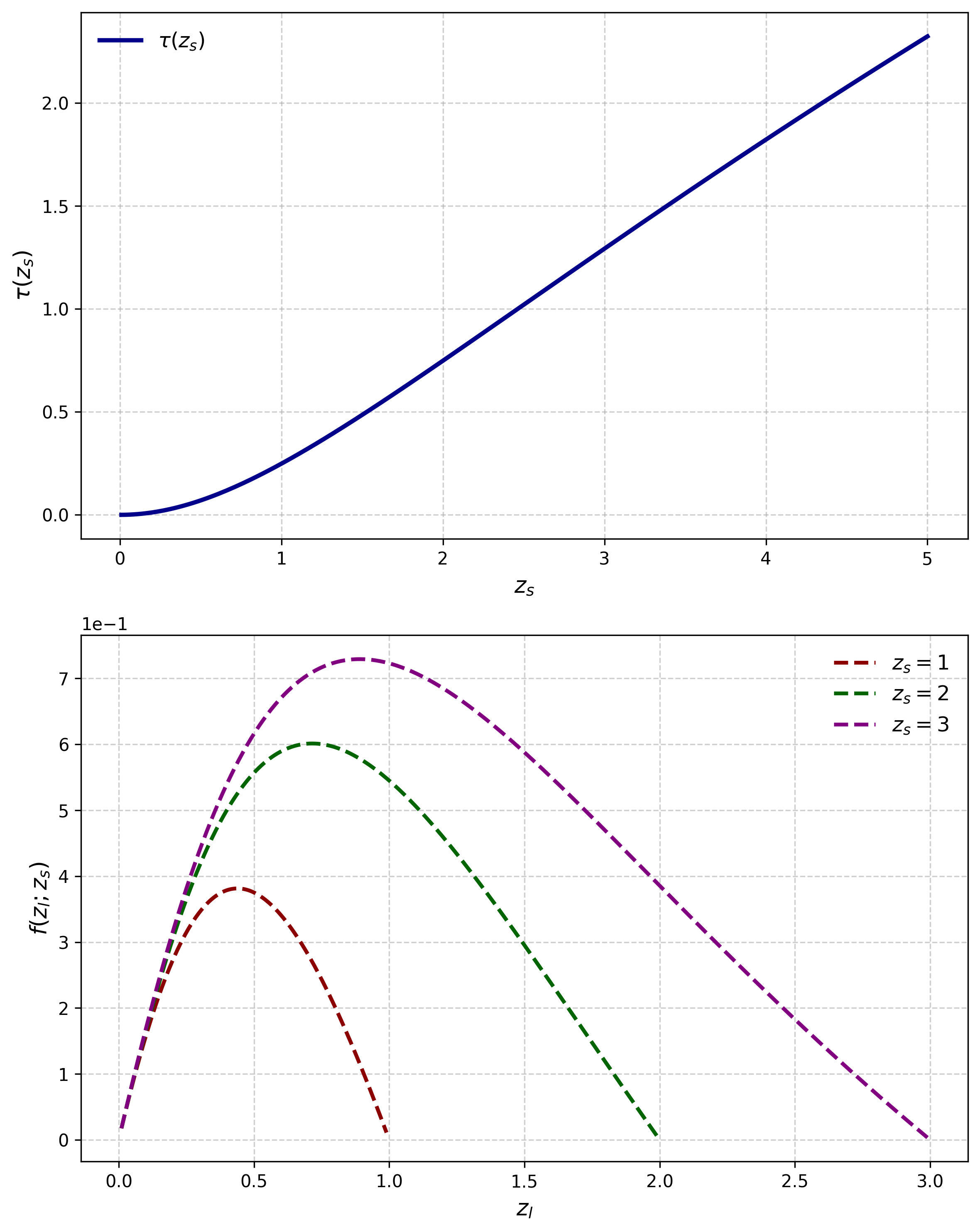}
    \caption{The upper and lower panels correspond to the optical depth for PBH lensing as a function of source redshift and to its probability density function (PDF) as a function of lens redshift, respectively. As shown, the PDF curve first increases and then decreases, exhibiting a distinct peak. This behavior indicates that our assumption in Eq.~\ref{defpro} is well justified in this context.}
    \label{fig:pbh_lensing}
\end{figure}

The corresponding probability density function (PDF) takes the form:
\begin{equation}
    \label{point_mass}
    f(z_l;z_s)=\frac{27}{2}f_{\mathrm{PBH}}\Omega_{\mathrm{DM}}\,\frac{H_0^2}{cH(z_l)}\frac{D_{\mathrm{l}}D_{\mathrm{ls}}}{D_{\mathrm{s}}}(1+z_l)^2.
\end{equation}
The results are shown in Fig.~\ref{fig:pbh_lensing}.

We next consider the form of the amplification factor, the amplification factor for a point-mass lens is expressed as:
\begin{equation}
\begin{aligned}
    \label{point_mass_af}
    F(f)=\exp & \left[\frac{\pi w}{4}+i \frac{w}{2}\left(\ln \left(\frac{w}{2}\right)-2 \phi_m(y)\right)\right] \\
    & \times \Gamma\left(1-\frac{i}{2} w\right)\,{}_1 F_1\left(\frac{i}{2} w, 1 ; \frac{i}{2} w y^2\right),
\end{aligned}
\end{equation}
where $w = 8\pi G M_{lz}f/c^3$, $\phi_m(y)=(x_m-y)^2/2-\ln x_m$, $x_m=(y+\sqrt{y^2+4})/2$, and ${}_1F_1$ denotes the confluent hypergeometric function, where $M_{lz}$ is the redshifted lens mass.  

Since the confluent hypergeometric function is computationally demanding, even with numerical optimizations, we approximate the averaged amplification factor in Eq.~\ref{averageamplitudefactor} by replacing $\left\langle |F(f;y)|^2 \right\rangle$ with $|F(f;\langle y\rangle)|^2$ for simplicity.  

\begin{figure}
    \centering
    \includegraphics[width=1.1\linewidth]{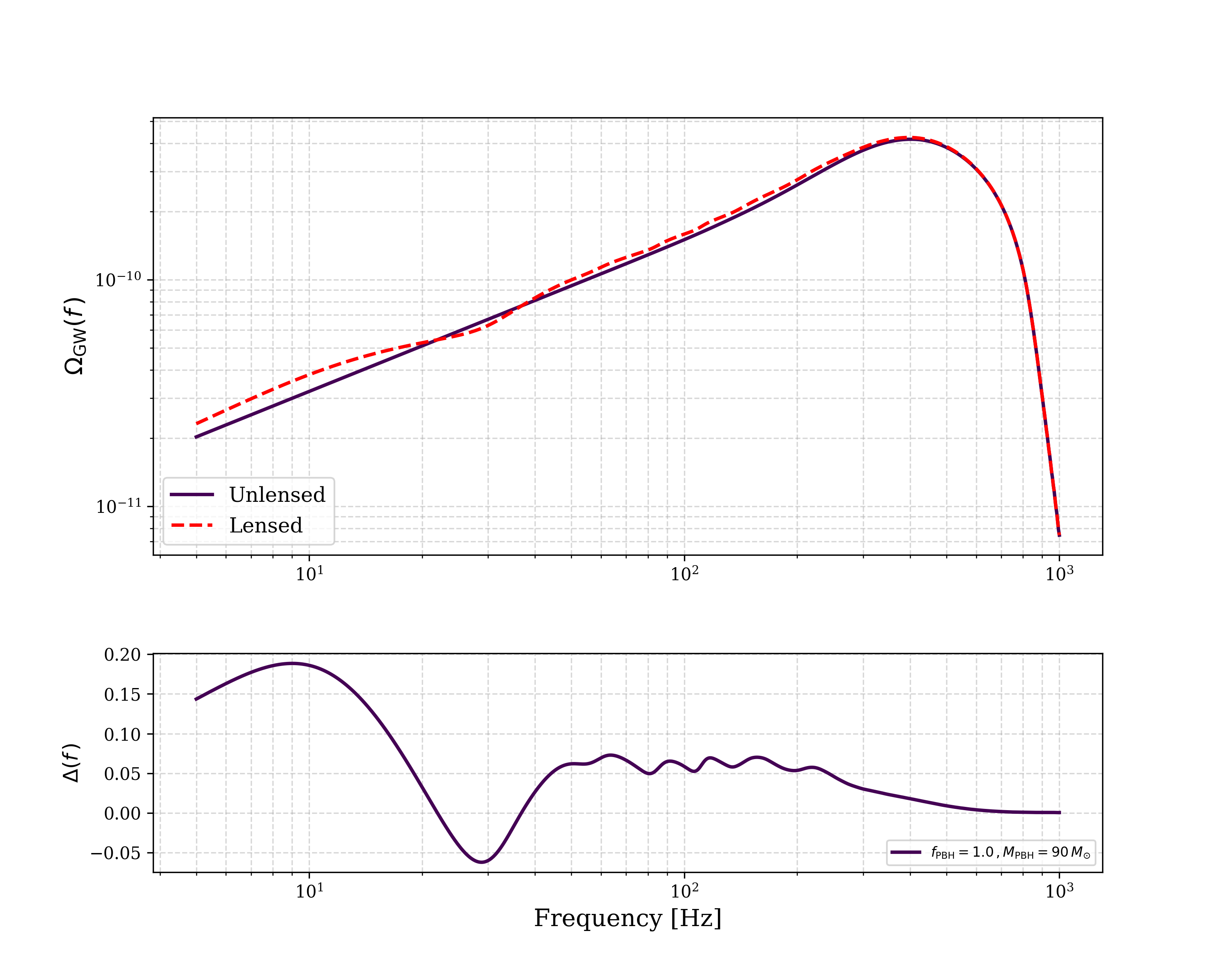}
    \caption{In the upper panel, the purple solid line and the red dashed line represent the stochastic gravitational-wave spectra without lensing and with lensing, respectively, where we adopt a uniform value of $t_{\min}=50\,\mathrm{Myr}$. The lower panel shows the difference between the lensed and unlensed spectra, quantified by Eq.~\ref{delta_f}. The adopted lensing parameters are $f_{\mathrm{PBH}}=1.0$ and $M_{\mathrm{PBH}}=90\,M_{\odot}$.}
    \label{lensedandunlensingpl}
\end{figure}

We show the lensed and unlensed energy spectra, together with their relative difference, in Fig.~\ref{lensedandunlensingpl}.

\section{Conclusion and Discussion}
In this section we will summary the all result we have got above, and discuss about how the gravitational lensing effect influence the energy spectrum of SGWB. 

For PBH lensing, the lower panel of Fig.~\ref{lensedandunlensingpl} shows that the relative difference between the lensed and unlensed spectra can reach the level of $10^{-1}$. This enhancement arises because we assume PBHs to be viable candidates for dark matter, thereby assigning them a widespread cosmological distribution. Consequently, the optical depth of PBH lensing is significantly large, making the lensing effect much more pronounced.

\begin{figure*}
    \centering
    \includegraphics[width=0.7\linewidth]{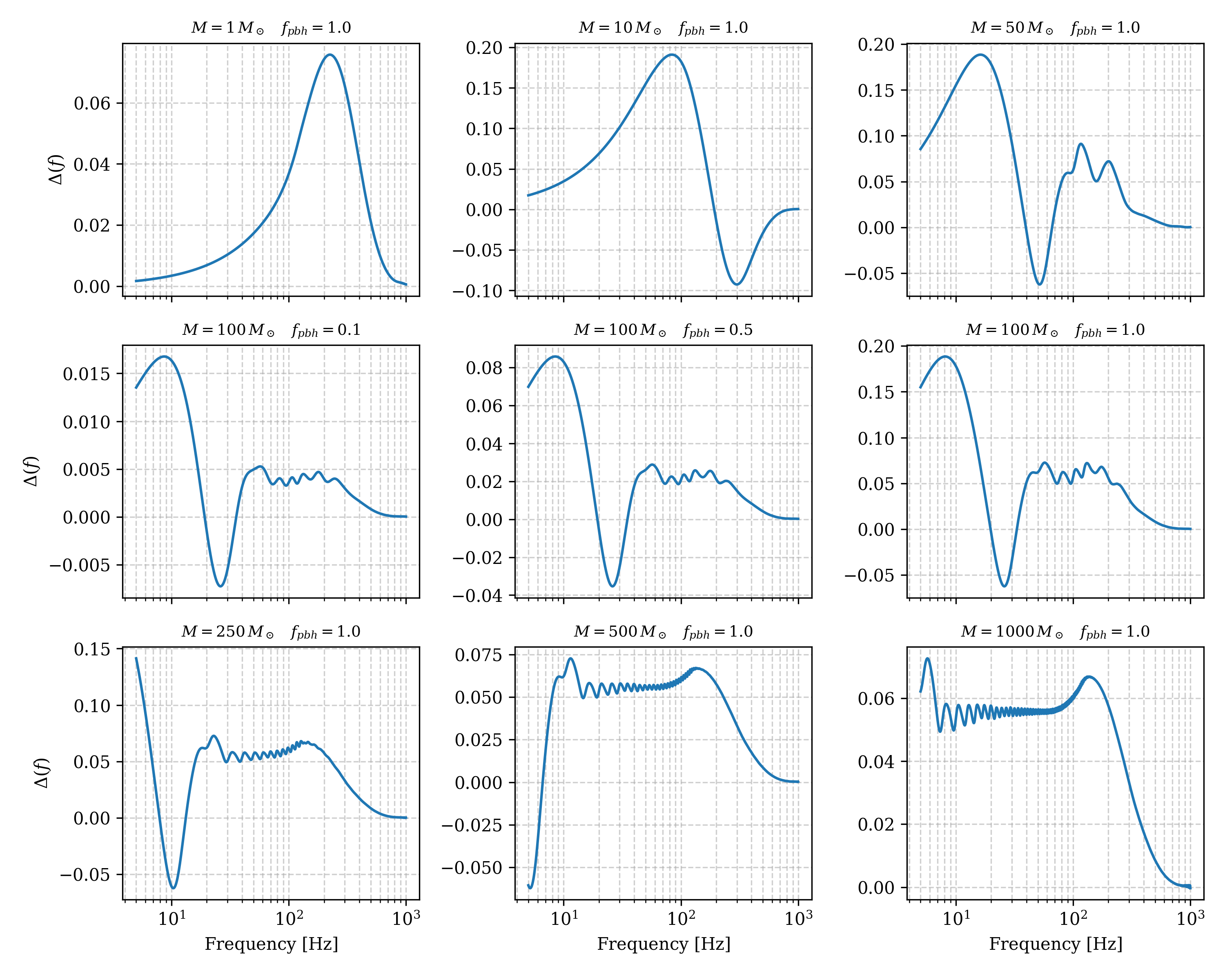}
    \caption{As shown in the figure, in the first and third row, we fix the PBH abundance at $f_{\mathrm{PBH}}=1$ and vary the lens mass as $M=1\,,10\,,50\,,250\,,500\,,1000\,M_{\odot}$, these two rows show how the lens mass influences the relative difference between the lensed and unlensed spectra. For the second row, we fix the PBH mass at $M_{\mathrm{PBH}}=100\,M_{\odot}$ and vary the PBH abundance as $f_{\mathrm{PBH}}=0.1\,,0.5\,,1.0$, this row shows how the PBH abundance affects the relative difference between the lensed and
unlensed spectra, with higher values of $f_{\mathrm{PBH}}$ leading to a more pronounced lensing effect.}
    \label{fig:total_relation}
\end{figure*}

In fact, primordial black hole (PBH) lensing is characterized by two fundamental parameters: the PBH abundance, $f_{\mathrm{PBH}}$, and the PBH mass, $M_{\mathrm{PBH}}$. Variations in these parameters can lead to distinct modifications of the intrinsic SGWB spectrum. To elucidate their respective influences, we present Fig.~\ref{fig:total_relation}, which systematically illustrates the dependence of the lensed SGWB on these parameters. From a physical perspective, $f_{\mathrm{PBH}}$ specifies the fraction of dark matter composed of PBHs and thus directly determines the lensing optical depth, whereas $M_{\mathrm{PBH}}$ sets the characteristic lensing scale and governs the frequency-dependent diffraction features. By isolating their roles, we can disentangle the respective contributions of lens abundance and lens mass to the modification of the SGWB spectrum.

\begin{figure*}
    \centering
    \includegraphics[width=0.8\linewidth]{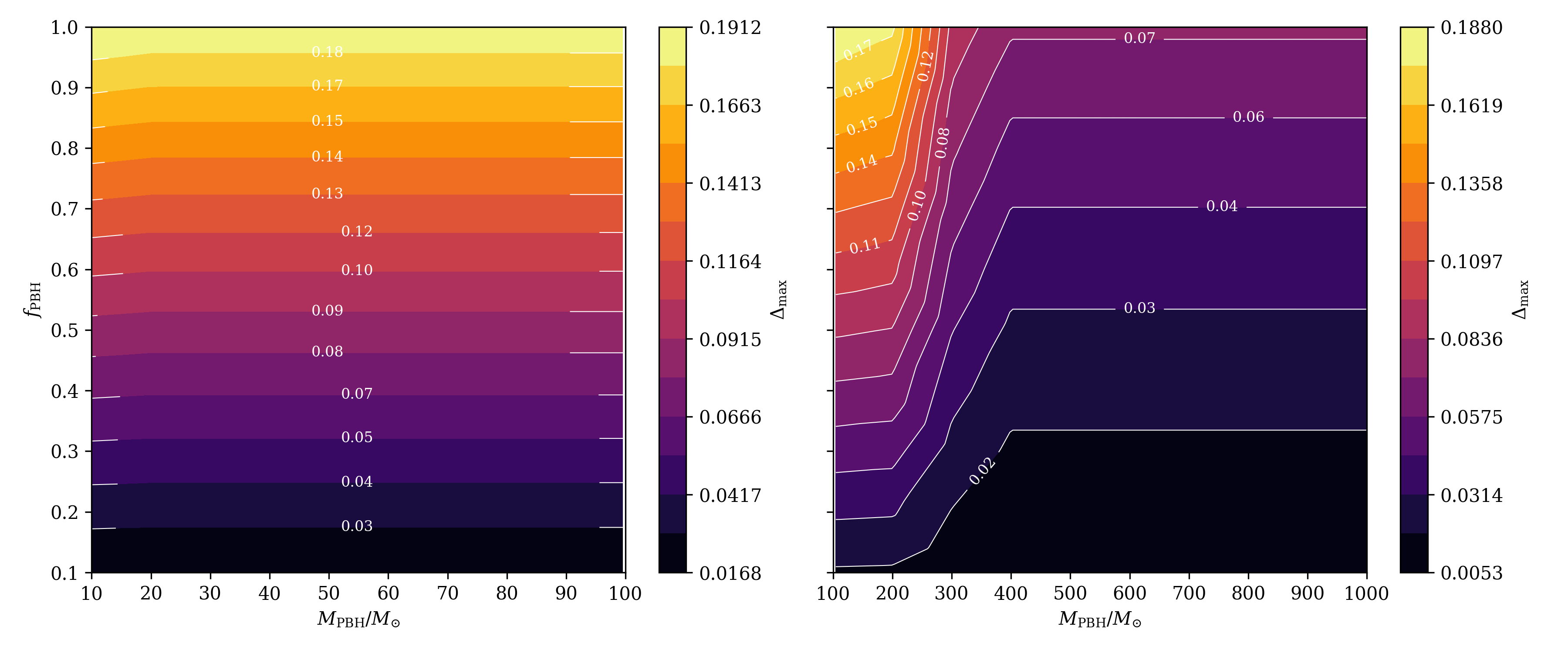}
    \caption{As shown in the figure, we plot the characteristic quantity $\Delta_{\mathrm{max}}$ as a function of the two lensing parameters, $f_{\mathrm{PBH}}$ and $M_{\mathrm{PBH}}$. The left panel corresponds to the mass range $10$–$100,M_{\odot}$, while the right panel covers $100$–$1000,M_{\odot}$. The color shading indicates the magnitude of $\Delta_{\mathrm{max}}$, while the overlaid contour lines mark regions of equal values, thereby highlighting the variation and local structure of $\Delta_{\mathrm{max}}$ across the parameter space.}
    \label{fig:peak_delta}    
\end{figure*}

\begin{figure*}
    \centering
    \includegraphics[width=0.8\linewidth]{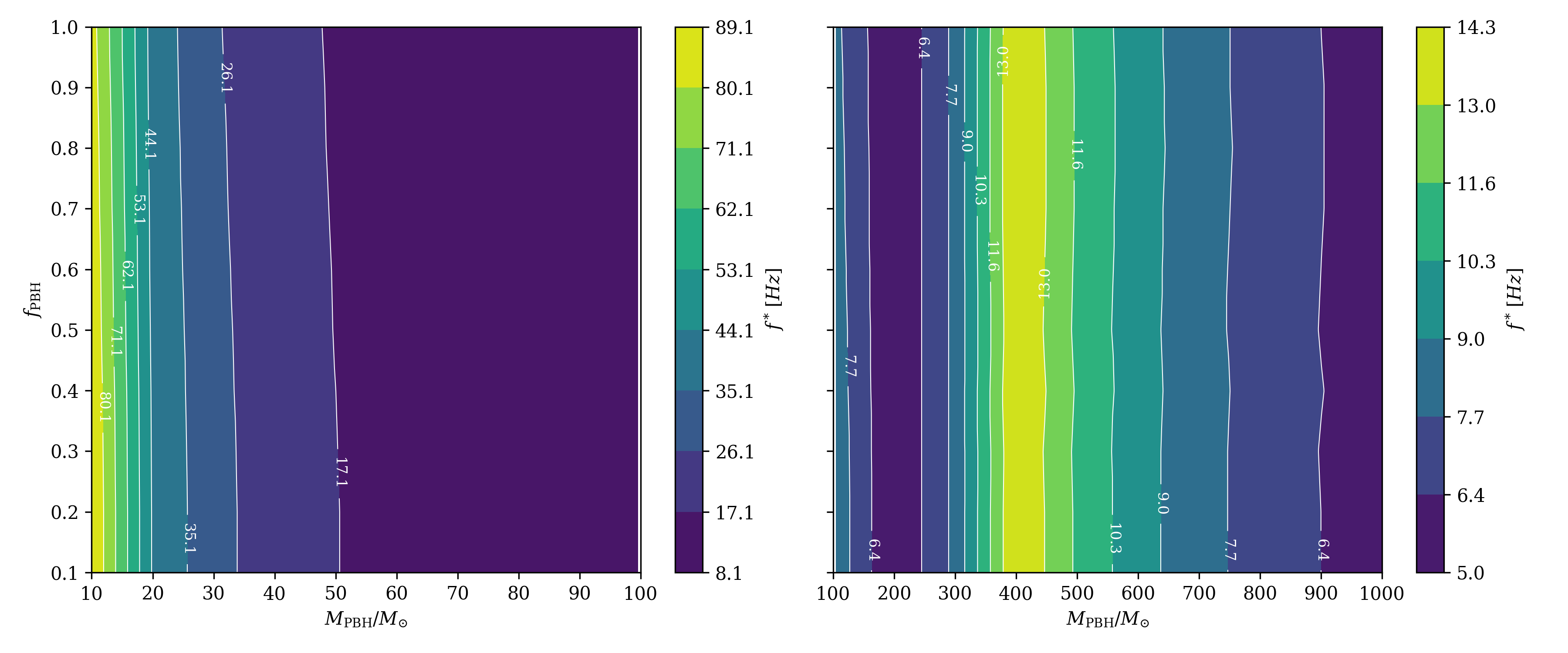}
    \caption{Consistent with the previous figure, this plot shows the characteristic quantity $f^{*}$ as a function of the two lensing parameters, $f_{\mathrm{PBH}}$ and $M_{\mathrm{PBH}}$.}
    \label{fig:peak_freq} 
\end{figure*}
As shown in Fig.~\ref{fig:total_relation}, the peak features of the relative difference between the lensed and unlensed spectra vary with different lensing parameters, both in terms of maximum amplitude and the corresponding frequency. To quantify these effects, we introduce two characteristic quantities, $\Delta_{\mathrm{max}}$ and $f^{*}$, which denote the peak amplitude and its associated frequency, respectively. These quantities provide a useful framework for probing lensing parameters once observational data of the SGWB become available.

From Fig.~\ref{fig:peak_delta}, it can be seen that in the mass range of $10$–$100\,M_{\odot}$, variations in $M_{\mathrm{PBH}}$ have only a minor impact on $\Delta_{\mathrm{max}}$, while $f_{\mathrm{PBH}}$ plays a dominant role in determining its value. In contrast, for higher masses in the range of $100$–$1000\,M_{\odot}$, $\Delta_{\mathrm{max}}$ decreases as $M_{\mathrm{PBH}}$ increases. Furthermore, as shown in Fig.~\ref{fig:peak_freq}, $f_{\mathrm{PBH}}$ has little influence on $f^{*}$, indicating that the PBH mass is the primary parameter governing the frequency-dependent shape of the SGWB spectrum.

Since the SGWB from binary black hole mergers has not yet been detected, there is currently no observational data available for further analysis, such as constraining dark matter parameters through comparisons between lensed and unlensed spectra. Nevertheless, based on analytical formulations, this study demonstrates that the lensed SGWB exhibits distinct and measurable features. First, the peak relative difference can reach approximately $20\%$, indicating that PBH lensing can induce a substantial modification to the SGWB spectrum. This peak amplitude shows weak dependence on the PBH mass, implying that it could serve as an effective probe of the PBH abundance. In contrast, the frequency at which the lensed spectrum attains its maximum is highly sensitive to the PBH mass, suggesting a complementary diagnostic of the lens properties. By jointly analyzing these two features, future detections of the SGWB may enable robust constraints on PBH-related dark matter scenarios. Furthermore, the overall shape of the lensed spectrum varies noticeably with different PBH parameters, providing a promising basis for model fitting and statistical inference once observational data become available.

\section*{Acknowledgments}
This work was supported by National Key Research and Development Program of China (No. 2024YFC2207400) and National Natural Science Foundation of China (No. 12222302).
\appendix*

\nocite{*}

\bibliography{used}% Produces the bibliography via BibTeX.

@article{phinney2001practical,
  title={A practical theorem on gravitational wave backgrounds},
  author={Phinney, ES},
  journal={arXiv preprint astro-ph/0108028},
  year={2001}
}

@ARTICLE{2025ChPhL..42e9201Z,
       author = {{Zhao}, Haoyu and {Zhang}, Yuanhao and {Fan}, Xilong and {Han}, Wenbiao},
        title = "{Gravitational Wave Background from Extreme-Mass-Ratio Inspirals}",
      journal = {Chinese Physics Letters},
         year = 2025,
        month = may,
       volume = {42},
       number = {5},
          eid = {059201},
        pages = {059201},
          doi = {10.1088/0256-307X/42/5/059201},
       adsurl = {https://ui.adsabs.harvard.edu/abs/2025ChPhL..42e9201Z}
}

@ARTICLE{2022ChPhL..39k9801L,
       author = {{Liao}, Kai and {Biesiada}, Marek and {Zhu}, Zong-Hong},
        title = "{Strongly Lensed Transient Sources: A Review}",
      journal = {Chinese Physics Letters},
         year = 2022,
        month = nov,
       volume = {39},
       number = {11},
          eid = {119801},
        pages = {119801},
          doi = {10.1088/0256-307X/39/11/119801},
archivePrefix = {arXiv},
       eprint = {2207.13489},
 primaryClass = {astro-ph.HE},
       adsurl = {https://ui.adsabs.harvard.edu/abs/2022ChPhL..39k9801L}
}

@ARTICLE{2017NatCo...8.1148L,
       author = {{Liao}, Kai and {Fan}, Xi-Long and {Ding}, Xuheng and {Biesiada}, Marek and {Zhu}, Zong-Hong},
        title = "{Precision cosmology from future lensed gravitational wave and electromagnetic signals}",
      journal = {Nature Communications},
         year = 2017,
        month = oct,
       volume = {8},
          eid = {1148},
        pages = {1148},
          doi = {10.1038/s41467-017-01152-9},
archivePrefix = {arXiv},
       eprint = {1703.04151},
 primaryClass = {astro-ph.CO},
       adsurl = {https://ui.adsabs.harvard.edu/abs/2017NatCo...8.1148L}
}

@article{zhu2013gravitational,
  title={On the gravitational wave background from compact binary coalescences in the band of ground-based interferometers},
  author={Zhu, Xing-Jiang and Howell, Eric J and Blair, David G and Zhu, Zong-Hong},
  journal={Monthly Notices of the Royal Astronomical Society},
  volume={431},
  number={1},
  pages={882--899},
  year={2013},
  publisher={The Royal Astronomical Society}
}

@article{aghanim2020planck,
  title={Planck 2018 results. VI. Cosmological parameters},
  author={Aghanim, N and others},
  journal={arXiv preprint arXiv:1807.06209},
  year={2020}
}

@article{regimbau2008astrophysical,
  title={Astrophysical sources of a stochastic gravitational-wave background},
  author={Regimbau, Tania and Mandic, Vuk},
  journal={Classical and Quantum Gravity},
  volume={25},
  number={18},
  pages={184018},
  year={2008},
  publisher={IOP Publishing}
}

@article{wu2012accessibility,
  title={Accessibility of the gravitational-wave background due to binary coalescences to second and third generation gravitational-wave detectors},
  author={Wu, Chengjiang and Mandic, Vuk and Regimbau, Tania},
  journal={Physical Review D—Particles, Fields, Gravitation, and Cosmology},
  volume={85},
  number={10},
  pages={104024},
  year={2012},
  publisher={APS}
}

@article{dominik2012double,
  title={Double compact objects. I. The significance of the common envelope on merger rates},
  author={Dominik, Michal and Belczynski, Krzysztof and Fryer, Christopher and Holz, Daniel E and Berti, Emanuele and Bulik, Tomasz and Mandel, Ilya and O'shaughnessy, Richard},
  journal={The Astrophysical Journal},
  volume={759},
  number={1},
  pages={52},
  year={2012},
  publisher={IOP Publishing}
}

@article{vangioni2015impact,
  title={The impact of star formation and gamma-ray burst rates at high redshift on cosmic chemical evolution and reionization},
  author={Vangioni, Elisabeth and Olive, Keith A and Prestegard, Tanner and Silk, Joseph and Petitjean, Patrick and Mandic, Vuk},
  journal={Monthly Notices of the Royal Astronomical Society},
  volume={447},
  number={3},
  pages={2575--2587},
  year={2015},
  publisher={Oxford University Press}
}

@article{abbott2021population,
  title={Population properties of compact objects from the second LIGO--Virgo gravitational-wave transient catalog},
  author={Abbott, Rich and Abbott, TD and Abraham, S and Acernese, Fausto and Ackley, K and Adams, A and Adams, C and Adhikari, RX and Adya, VB and Affeldt, Christoph and others},
  journal={The Astrophysical journal letters},
  volume={913},
  number={1},
  pages={L7},
  year={2021},
  publisher={IoP Publishing}
}

@article{hopkins2006normalization,
  title={On the normalization of the cosmic star formation history},
  author={Hopkins, Andrew M and Beacom, John F},
  journal={The Astrophysical Journal},
  volume={651},
  number={1},
  pages={142},
  year={2006},
  publisher={IOP Publishing}
}

@article{fardal2007evolutionary,
  title={On the evolutionary history of stars and their fossil mass and light},
  author={Fardal, Mark A and Katz, Neal and Weinberg, David H and Dav{\'e}, Romeel},
  journal={Monthly Notices of the Royal Astronomical Society},
  volume={379},
  number={3},
  pages={985--1002},
  year={2007},
  publisher={Blackwell Publishing Ltd Oxford, UK}
}

@article{wilkins2008evolution,
  title={The evolution of stellar mass and the implied star formation history},
  author={Wilkins, Stephen M and Trentham, Neil and Hopkins, Andrew M},
  journal={Monthly Notices of the Royal Astronomical Society},
  volume={385},
  number={2},
  pages={687--694},
  year={2008},
  publisher={The Royal Astronomical Society}
}

@inproceedings{thorne1997gravitational,
  title={Gravitational Radiation-a New Window onto the Universe.(Karl Schwarzschild Lecture 1996)},
  author={Thorne, Kip S},
  booktitle={Reviews in Modern Astronomy, v. 10,(1997), p. 1-28.},
  volume={10},
  pages={1--28},
  year={1997}
}

@article{ajith2008template,
  title={Template bank for gravitational waveforms from coalescing binary black holes: Nonspinning binaries},
  author={Ajith, Parameswaran and Babak, Stanislav and Chen, Yanbei and Hewitson, Martin and Krishnan, Badri and Sintes, AM and Whelan, John T and Bruegmann, Bernd and Diener, P and Dorband, Nils and others},
  journal={Physical Review D—Particles, Fields, Gravitation, and Cosmology},
  volume={77},
  number={10},
  pages={104017},
  year={2008},
  publisher={APS}
}

@article{baraldo1999gravitationally,
  title={Gravitationally induced interference of gravitational waves by a rotating massive object},
  author={Baraldo, Christian and Hosoya, Akio and Nakamura, Takahiro T},
  journal={Physical Review D},
  volume={59},
  number={8},
  pages={083001},
  year={1999},
  publisher={APS}
}

@article{takahashi2003wave,
  title={Wave effects in the gravitational lensing of gravitational waves from chirping binaries},
  author={Takahashi, Ryuichi and Nakamura, Takashi},
  journal={The Astrophysical Journal},
  volume={595},
  number={2},
  pages={1039},
  year={2003},
  publisher={IOP Publishing}
}

@article{zhou2022search,
  title={Search for lensing signatures from the latest fast radio burst observations and constraints on the abundance of primordial black holes},
  author={Zhou, Huan and Li, Zhengxiang and Liao, Kai and Niu, Chenhui and Gao, He and Huang, Zhiqi and Huang, Lu and Zhang, Bing},
  journal={The Astrophysical Journal},
  volume={928},
  number={2},
  pages={124},
  year={2022},
  publisher={IOP Publishing}
}

@article{liao2019wave,
  title={The wave nature of continuous gravitational waves from microlensing},
  author={Liao, Kai and Biesiada, Marek and Fan, Xi-Long},
  journal={The Astrophysical Journal},
  volume={875},
  number={2},
  pages={139},
  year={2019},
  publisher={IOP Publishing}
}

@article{grishchuk1977gravitational,
  title={Gravitational waves in the cosmos and the laboratory},
  author={Grishchuk, Leonid P},
  journal={Soviet Physics Uspekhi},
  volume={20},
  number={4},
  pages={319},
  year={1977},
  publisher={IOP Publishing}
}

@article{regimbau2011astrophysical,
  title={The astrophysical gravitational wave stochastic background},
  author={Regimbau, Tania},
  journal={Research in Astronomy and Astrophysics},
  volume={11},
  number={4},
  pages={369},
  year={2011},
  publisher={IOP Publishing}
}

@article{abbott2018kagra,
  title={KAGRA and LIGO Scientific and VIRGO Collaborations},
  author={Abbott, BP and Abbott, R and Abbott, TD and Abraham, S and Acernese, F and Ackley, K and Adams, C and Adya, VB and Affeldt, C and Agathos, M and others},
  journal={Living Rev. Rel},
  volume={21},
  number={3},
  year={2018}
}

@article{zhu2011stochastic,
  title={Stochastic gravitational wave background from coalescing binary black holes},
  author={Zhu, Xing-Jiang and Howell, Eric and Regimbau, Tania and Blair, David and Zhu, Zong-Hong},
  journal={The Astrophysical Journal},
  volume={739},
  number={2},
  pages={86},
  year={2011},
  publisher={IOP Publishing}
}

@article{oguri2018effect,
  title={Effect of gravitational lensing on the distribution of gravitational waves from distant binary black hole mergers},
  author={Oguri, Masamune},
  journal={Monthly Notices of the Royal Astronomical Society},
  volume={480},
  number={3},
  pages={3842--3855},
  year={2018},
  publisher={Oxford University Press}
}

@article{bliokh1975diffraction,
  title={Diffraction of light and lens effect of the stellar gravitation field},
  author={Bliokh, PV and Minakov, AA},
  journal={Astrophysics and Space Science},
  volume={34},
  number={2},
  pages={L7--L9},
  year={1975},
  publisher={Springer}
}

@ARTICLE{1998PhRvL..80.1138N,
       author = {{Nakamura}, Takahiro T.},
        title = "{Gravitational Lensing of Gravitational Waves from Inspiraling Binaries by a Point Mass Lens}",
      journal = {\prl},
         year = 1998,
        month = feb,
       volume = {80},
       number = {6},
        pages = {1138-1141},
}

@article{takahashi2017arrival,
  title={Arrival time differences between gravitational waves and electromagnetic signals due to gravitational lensing},
  author={Takahashi, Ryuichi},
  journal={The Astrophysical Journal},
  volume={835},
  number={1},
  pages={103},
  year={2017},
  publisher={IOP Publishing}
}

@article{hanson2010weak,
  title={Weak lensing of the CMB},
  author={Hanson, Duncan and Challinor, Anthony and Lewis, Antony},
  journal={General Relativity and Gravitation},
  volume={42},
  number={9},
  pages={2197--2218},
  year={2010},
  publisher={Springer}
}

@article{bertone2018history,
  title={History of dark matter},
  author={Bertone, Gianfranco and Hooper, Dan},
  journal={Reviews of Modern Physics},
  volume={90},
  number={4},
  pages={045002},
  year={2018},
  publisher={APS}
}

@article{sasaki2018primordial,
  title={Primordial black holes—perspectives in gravitational wave astronomy},
  author={Sasaki, Misao and Suyama, Teruaki and Tanaka, Takahiro and Yokoyama, Shuichiro},
  journal={Classical and Quantum Gravity},
  volume={35},
  number={6},
  pages={063001},
  year={2018},
  publisher={IOP Publishing}
}

@article{green2021primordial,
  title={Primordial Black Holes as a dark matter candidate},
  author={Green, Anne M and Kavanagh, Bradley J},
  journal={Journal of Physics G: Nuclear and Particle Physics},
  volume={48},
  number={4},
  pages={043001},
  year={2021},
  publisher={IOP Publishing}
}

@article{frampton2016primordial,
  title={The primordial black hole mass range},
  author={Frampton, Paul H},
  journal={Modern Physics Letters A},
  volume={31},
  number={12},
  pages={1650064},
  year={2016},
  publisher={World Scientific}
}

@article{liao2020probing,
  title={Probing compact dark matter with gravitational wave fringes detected by the Einstein Telescope},
  author={Liao, Kai and Tian, Shuxun and Ding, Xuheng},
  journal={Monthly Notices of the Royal Astronomical Society},
  volume={495},
  number={2},
  pages={2002--2006},
  year={2020},
  publisher={Oxford University Press}
}

@article{urrutia2022lensing,
  title={Lensing of gravitational waves as a probe of compact dark matter},
  author={Urrutia, Juan and Vaskonen, Ville},
  journal={Monthly Notices of the Royal Astronomical Society},
  volume={509},
  number={1},
  pages={1358--1365},
  year={2022},
  publisher={Oxford University Press}
}

@article{ji2018strong,
  title={Strong lensing of gamma ray bursts as a probe of compact dark matter},
  author={Ji, Lingyuan and Kovetz, Ely D and Kamionkowski, Marc},
  journal={Physical Review D},
  volume={98},
  number={12},
  pages={123523},
  year={2018},
  publisher={APS}
}

@article{halder2021bounds,
  title={Bounds on abundance of primordial black hole and dark matter from EDGES 21-cm signal},
  author={Halder, Ashadul and Banerjee, Shibaji},
  journal={Physical Review D},
  volume={103},
  number={6},
  pages={063044},
  year={2021},
  publisher={APS}
}

\end{document}